\documentclass[showpacs,showkeys,amsmath,amssymb,preprint]{revtex4}
\usepackage{epsfig,dcolumn,bm}

\begin{document}

\title{Low-lying $ud{\bar s}{\bar s}$ configurations in a non-relativistic constituent quark model}

\author{W.L. Wang$^1$}
\email{wlwang@ihep.ac.cn}
\author{F. Huang$^2$}
\author{Z.Y. Zhang$^3$}
\author{Y.W. Yu$^3$}
\author{F. Liu$^1$}
\affiliation
{\small $^1$Institute of Particle Physics, Huazhong Normal University, Wuhan 430079, China\\
$^2$CCAST (World Laboratory), P.O. Box 8730, Beijing 100080, China \\
$^3$Institute of High Energy Physics, P.O. Box 918-4, Beijing 100049, China}

\begin{abstract}
The energies of the low-lying isoscalar and isovector $ud\bar s\bar
s$ configurations with spin-parity $J^P=0^+$, $1^+$, and $2^+$ are
calculated in a non-relativistic constituent quark model by use of
the variational method. The contributions of various parts of the
quark-quark interacting potentials including the $s$-channel
interaction are investigated, and the effect of different forms of
confinement potential is examined. The model parameters are
determined by the same method as in our previous work, and they
still can satisfactorily describe the nucleon-nucleon scattering
phase shifts and the hyperon-nucleon cross sections. The parameters
of the $s$-channel interaction are fixed by the masses of $K$ and
$K^*$ mesons, for which the size parameter is taken to be two
possible values. When it is chosen as the same as baryons', the
numerical results show that the masses of all the $ud\bar s\bar s$
configurations are higher than the corresponding meson-meson
thresholds. But when the size parameter for the $K$ and $K^*$ mesons
is adjusted to be smaller than that for the baryons, the $ud\bar
s\bar s$ configuration with $I=0$ and $J^P=1^+$ is found to lie
lower than the $K^*K^*$ threshold, furthermore, this state has a
very small $KK^*$ component and the interaction matrix elements
between this state and $KK^*$ is comparatively small, thus its
coupling to the $KK^*$ channel will consequently be weak and it
might be regarded as a possible tetraquark candidate.
\end{abstract}



\maketitle

\section{Introduction}

So far, all the observed hadrons can be classified into two types,
i.e. ``baryons" composed of $qqq$ and ``mesons" composed of $q\bar
q$. But in principle, the QCD fundamental theory doesn't exclude the
existence of the states containing more than three quarks, i.e. the
so-called multi-quark states. Since Jaffe predicted the $H$ particle
($uuddss$) in 1977 \cite{jaffe}, the research on multi-quark states
has always been an attractive topic for nearly three decades in both
theoretical and experimental studies. But up to now, there is no
convincing evidence of their existence in experiments. In 2003, LEPS
Collaboration reported the possible existence of the $\Theta^+$
pentaquark \cite{tnak03}, and subsequently some laboratories also
reported the similar results \cite{5q-review}. At the same time,
several laboratories reported the negative results \cite{5q-review}.
Although its existence is still questioned, the $\Theta^+$ particle
has motivated a number of theoretical and experimental studies of
pentaquarks and further the multi-quark states.

Besides dibaryon and pentaquark, the possible $ud\bar s\bar s$
tetraquark is another interesting multi-quark system, and many works
have been devoted to the investigation of this state in the past few
years
\cite{jaffe1977,vijande04,burns05,karliner05,kanada05,cui06,chen06}.
In Ref. \cite{jaffe1977}, using the MIT bag model, Jaffe performed a
wide study for the spectrum of the $4q$ states, and he predicted the
masses of the isovector $J^P=0^+$ and isoscalar $J^P=1^+$ $ud\bar
s\bar s$ states to be 1.55 GeV and 1.65 GeV, respectively. In Ref.
\cite{vijande04}, Vijande {\it et al.} analyzed the $ud\bar s\bar s$
systems for both isospin $I=0$ and $I=1$ channels in a
constituent-quark model, and they didn't find any stable $ud\bar
s\bar s$ tetraquark state. In Ref. \cite{burns05}, Burns {\it et
al.} claimed that the $\Theta^+$ particle suggest there should exist
a $ud\bar s\bar s$ tetraquark state with $J^P=1^-$ and mass around
1.6 GeV. This state has a strong color-magnetic attraction and
decays into $KK$ channel via $P$-wave with a width around $10-100$
MeV. In Ref. \cite{karliner05}, Karliner and Lipkin argued that the
isoscalar $J^P=0^+$ $ud\bar s\bar s$ tetraquark state is a cousin of
the $\Theta^+$ pentaquark, and for this state the lowest allowed
decay mode is a four-body $KK\pi\pi$ channel with a very small phase
space and a distinctive experimental signature. However, we should
note that from the non-clustered quark degree of freedom, the
isoscalar $J^P=0^+$ $ud\bar s\bar s$ state within the spatially
symmetric configuration is not allowed due to the Pauli principle.
In Ref. \cite{kanada05}, Kanada-En'yo {\it et al.} studied the
$ud\bar s\bar s$ system in the framework of the flux-tube quark
model, and they pointed that the isoscalar $J^P=1^+$ $ud\bar s\bar
s$ is a stable and low-lying tetraquark state with a mass around 1.4
GeV. It can decay into $K^*K$ via $S$-wave with its width around
20-80 MeV, while the isovector $J^P=0^+$ $ud\bar s\bar s$ tetraquark
does not exist. Recently, Cui {\it et al.} \cite{cui06} calculated
the masses of the isoscalar $J^P=1^+$ $ud\bar s\bar s$ systems by
using the color-magnetic interaction Hamiltonian with SU(3) flavor
symmetry breaking, and they found the $ud\bar s\bar s$ tetraquark
lies around 1347 MeV with a narrow width since it can not decay into
the $KK^*$ channel. In Ref. \cite{chen06}, Chen {\it et al.}
performed a QCD sum rule study for the isovector $J^P=0^+$ $ud\bar
s\bar s$ system, and they obtained a tetraquark state with a mass
around 1.5 GeV.

To sum up, the existence and properties of the possible $ud\bar
s\bar s$ tetraquark state are presently model dependent. Further
theoretical and experimental investigations of this state via
different approaches seem to be significant and essential.

It is a general consensus that QCD is the underlying theory of the
strong interaction. However, as the non-perturbative QCD effect is
very important for light quark systems in the low energy region and
it is difficult to be seriously solved, people still need
QCD-inspired models to be a bridge connecting the QCD fundamental
theory and the experimental observations. In the past few years, we
have developed a non-relativistic constituent quark model, which has
been quite successful in reproducing the energies of the baryon
ground states, the binding energy of the deuteron, the
nucleon-nucleon ($NN$) scattering phase shifts, and the
hyperon-nucleon ($YN$) cross sections \cite{zyzhang97}. In this
model, the quark-quark interaction contains confinement, one-gluon
exchange (OGE) and boson exchanges stemming from scalar and
pseudoscalar nonets, with the boson exchange potentials deduced from
a linear interacting Lagrangian which is invariable under the chiral
transformation \cite{fhuang04kn}, and the short range quark-quark
interaction is provided by OGE and quark exchange effects.

Actually it is still a controversial problem in the low-energy
hadron physics whether gluon or Goldstone boson is the proper
effective degree of freedom besides the constituent quark. Glozman
and Riska proposed that the Goldstone boson is the only other proper
effective degree of freedom \cite{glozman96,glozman00}. But Isgur
gave a critique of the boson exchange model and insisted that the
OGE governs the baryon structure \cite{isgur001,isgur002}. Anyway,
it is still an open problem in the low-energy hadron physics whether
OGE or vector-meson exchange is the right mechanism  for describing
the short-range quark-quark interaction, or both of them are
important. Thus in Ref. \cite{lrdai03} we further extended our
original constituent quark model to include the coupling of the
quark and vector meson fields. The OGE that plays an important role
in the short-range quark-quark interaction in our original model is
now nearly replaced by the vector boson exchanges. This model has
also been successful in reproducing the energies of the baryon
ground states, the binding energy of the deuteron, and the
nucleon-nucleon scattering phase shifts \cite{lrdai03}.

Recently, we have extended our constituent quark model from the
study of baryon-baryon scattering processes to the baryon-meson
systems and the pentaquark $\Theta^+$ state
\cite{fhuang04nkdk,fhuang05lksk,fhuang05dklksk,fhuang04kn,fhuang05kne,fhuang06nphi,
fhuang04penta,dzhang05}. We found that some results are similar to
those given by the chiral unitary approach study, such as that both
the $\Delta K$ system with isospin $I=1$ and the $\Sigma K$ system
with $I=1/2$ have quite strong attractions
\cite{fhuang04nkdk,fhuang05lksk,fhuang05dklksk}. In the study of the
$KN$ scattering \cite{fhuang04nkdk,fhuang04kn,fhuang05kne}, we get a
considerable improvement not only on the signs but also on the
magnitudes of the theoretical phase shifts comparing with other's
previous work. We also studied the structures of the $\Theta^+$
particle \cite{fhuang04penta,dzhang05}, and found that the
theoretical mass of $\Theta^+$ is much higher than the experimental
value, thus we concluded that either the $\Theta^+$ particle does
not exist or it can not be explained by a five-quark cluster, which
is consistent with the high-statistic experimental negative results.

In this paper, using the variational method, we study the structures
of the isoscalar and isovector $ud\bar s\bar s$ configurations with
spin-parity $J^P=0^+$, $1^+$, and $2^+$ in our non-relativistic
constituent quark model with meson and one gluon exchanges included.
The contributions of various parts of the quark-quark interacting
potentials including the $s$-channel interaction are investigated.
The effect of different forms of confinement potential is examined,
and the configuration mixing between the states with same quantum
numbers is considered. The model parameters are determined by the
same method as in our previous work
\cite{fhuang05lksk,fhuang05dklksk,fhuang06nphi}, and they still can
satisfactorily describe the nucleon-nucleon scattering phase shifts
and the hyperon-nucleon cross sections. The $s$-channel
quark-antiquark interaction is complicated, and we take the zero
momentum approximation and fix the corresponding parameters by
fitting the masses of $K$ and $K^*$ mesons. With the size parameter
of $K$ and $K^*$ taken to be the same as baryons', the numerical
results show that the masses of all the $ud\bar s\bar s$
configurations are higher than the corresponding meson-meson
thresholds. But when the size parameter for the $K$ and $K^*$ mesons
is adjusted to be smaller than that for the baryons, the $ud\bar
s\bar s$ configuration with $I=0$ and $J^P=1^+$ is found to lie
lower than the $K^*K^*$ threshold, furthermore, this state has a
very small $KK^*$ component and the interaction matrix elements
between this state and $KK^*$ is comparatively small, thus its
coupling to the $KK^*$ channel will consequently be weak and it
might be regarded as a possible tetraquark candidate.

The paper is organized as follows. In the next section the framework
of the non-relativistic constituent quark model we used and the wave
functions of the $ud\bar s\bar s$ configurations are briefly
introduced. The calculated energies of six $ud\bar s\bar s$ states
are shown in Sec. III, where some discussion is presented as well.
Finally, the summary is given in Sec. IV.

\section{Formulation}

\subsection{Model}

Our non-relativistic constituent quark model has been widely
described in the literature
\cite{fhuang04nkdk,fhuang05lksk,fhuang04kn,fhuang05kne}, and the
details can be found in these references. Here we just give the
salient features of our model.

The total Hamiltonian of the $ud\bar s\bar s$ systems in the model
can be written as
\begin{equation}
H=\sum_{i}T_{i}-T_{G}+V_{12}+V_{\bar 3\bar 4}+\sum_{i=1,2\atop
j=3,4}V_{i\bar j},
\end{equation}
where $T_G$ is the kinetic energy operator for the center-of-mass
motion, and $V_{12}$, $V_{\bar 3\bar 4}$ and $V_{i\bar j}$ represent
the quark-quark, antiquark-antiquark and quark-antiquark
interactions, respectively,
\begin{equation}
V_{12}= V^{\rm OGE}_{12} + V^{\rm conf}_{12} + V^{\rm ch}_{12},
\end{equation}
where $V_{12}^{\rm OGE}$ is the OGE interaction,
\begin{eqnarray}
V^{\rm
OGE}_{12}=\frac{1}{4}g_{1}g_{2}\left({\bm\lambda}^c_1\cdot{\bm\lambda}^c_2\right)
\left\{\frac{1}{r_{12}}-\frac{\pi}{2} \delta({\bm r}_{12})
\left[\frac{1}{m^2_1}+\frac{1}{m^2_2}+\frac{4}{3}\frac{1}{m_1m_2}
({\bm \sigma}_1 \cdot {\bm \sigma}_2)\right]\right\},
\end{eqnarray}
and the confinement potential $V_{12}^{\rm conf}$, instead of the
quadratic form as used in our previous work, is taken as the linear
one,
\begin{eqnarray}
V_{12}^{\rm conf}=-\left({\bm \lambda}_{1}^{c}\cdot{{\bm
\lambda}_{2}^{c}}\right)\left(a_{12}^{c}r_{12} +a_{12}^{c0}\right).
\end{eqnarray}
$V^{\rm ch}_{12}$ represents the effective quark-quark potential
induced by the one-boson exchanges. In our original constituent
quark model, $V^{\rm ch}_{12}$ includes the scalar boson exchanges
and the pseudoscalar boson exchanges,
\begin{eqnarray}
V^{\rm ch}_{12} = \sum_{a=0}^8 V_{\sigma_a}({\bm
r}_{ij})+\sum_{a=0}^8 V_{\pi_a}({\bm r}_{ij}),
\end{eqnarray}
and when the model is extended to include the vector boson
exchanges, $V^{\rm ch}_{12}$ can be written as
\begin{eqnarray}
V^{\rm ch}_{12} = \sum_{a=0}^8 V_{\sigma_a}({\bm
r}_{ij})+\sum_{a=0}^8 V_{\pi_a}({\bm r}_{ij})+\sum_{a=0}^8
V_{\rho_a}({\bm r}_{ij}).
\end{eqnarray}
Here $\sigma_{0},...,\sigma_{8}$ are the scalar nonet fields,
$\pi_{0},..,\pi_{8}$ the pseudoscalar nonet fields, and
$\rho_{0},..,\rho_{8}$ the vector nonet fields. The expressions of
these potentials can be found in the literature
\cite{fhuang04nkdk,fhuang05lksk,fhuang04kn,fhuang05kne}.

$V_{\bar 3\bar 4}$ in Eq. (1) represents the antiquark-antiquark
interaction,
\begin{equation}
V_{\bar 3\bar 4}= V^{\rm OGE}_{\bar 3\bar 4} + V^{\rm conf}_{\bar
3\bar 4} + V^{\rm ch}_{\bar 3\bar 4},
\end{equation}
where $V^{\rm OGE}_{\bar 3\bar 4}$ and $V^{\rm conf}_{\bar 3\bar 4}$
can be obtained by replacing the
${\bm\lambda}^c_1\cdot{\bm\lambda}^c_2$ in Eqs. (3) and (4) with
${\bm\lambda}^{c*}_{\bar 3}\cdot{\bm\lambda}^{c*}_{\bar 4}$, and
$V^{\rm ch}_{\bar 3\bar 4}$ has the same form as $V^{\rm ch}_{12}$.

$V_{i\bar j}$ in Eq. (1) represents the quark-antiquark interaction,
\begin{equation}
V_{i\bar j}= V^{\rm OGE}_{i\bar j} + V^{\rm conf}_{i\bar j} + V^{\rm
ch}_{i\bar j} + V^{\rm s}_{i\bar j},
\end{equation}
where $V^{\rm OGE}_{i\bar j}$ and $V^{\rm conf}_{i\bar j}$ can be
obtained by replacing the ${\bm\lambda}^c_1\cdot{\bm\lambda}^c_2$ in
Eqs. (3) and (4) with
$-{\bm\lambda}^{c}_{i}\cdot{\bm\lambda}^{c*}_{\bar j}$, and
$V_{i\bar{j}}^{\rm ch}$ can be obtained from the G parity
transformation:
\begin{equation}
V_{i\bar{j}}^{\rm ch}=\sum_{k}(-1)^{G_k}V_{ij}^{{\rm ch},k},
\end{equation}
with $(-1)^{G_k}$ being the G parity of the $k$th meson.
$V^{s}_{i\bar j}$ denotes the $s$-channel quark-antiquark
interaction. For the $ud\bar s\bar s$ system, the $s$-channel
interaction includes $K$ and $K^*$ exchanges,
\begin{equation}
V_{i\bar j}^{s}=V_{s}^{K}+V_{s}^{K^*}.
\end{equation}
Taking the zero momentum approximation, the spatial part of
$V^{s}_{i\bar j}$ can be expressed as a delta-function. To flatten
the delta-function, we replace it by the Yukawa function, then
$V_{s}^{K}$ and $V_{s}^{K^*}$ can be expressed as:
\begin{eqnarray}
\label{vk} V_{s}^{K}=C^K\left(\frac{1-{\bm \sigma}_q \cdot {\bm
\sigma}_{\bar{q}}}{2}\right)_{s}\left(\frac{2 + 3\lambda_q \cdot
\lambda^*_{\bar{q}}}{6}\right)_{c}
\left(\frac{}{}2\frac{}{}\right)_{f}\frac{\Lambda^2}{r}e^{-\Lambda
r},
\end{eqnarray}
and
\begin{eqnarray}
\label{vks} V_{s}^{K^*}=C^{K^*}\left(\frac{3+{\bm \sigma}_q \cdot
{\bm \sigma}_{\bar{q}}}{2}\right)_{s}\left(\frac{2 + 3\lambda_q
\cdot \lambda^*_{\bar{q}}}{6}\right)_{c}
\left(\frac{}{}2\frac{}{}\right)_{f}\frac{\Lambda^2}{r}e^{-\Lambda
r},
\end{eqnarray}
where $C^K$ and $C^{K^*}$ are treated as parameters and we adjust
them to fit the masses of $K$ and $K^*$ mesons.

\subsection{Parameters}

{\small
\begin{table}[htb]
\caption{\label{para} Model parameters. The meson masses and the
cutoff masses: $m_{\sigma'}=980$ MeV, $m_{\kappa}=980$ MeV,
$m_{\epsilon}=980$ MeV, $m_{\pi}=138$ MeV, $m_K=495$ MeV,
$m_{\eta}=549$ MeV, $m_{\eta'}=957$ MeV, $m_{\rho}=770$ MeV,
$m_{K^*}=892$ MeV, $m_{\omega}=782$ MeV, $m_{\phi}=1020$ MeV, and
$\Lambda=1100$ MeV.}
\begin{center}
\begin{tabular*}{160mm}{@{\extracolsep\fill}cccc}
\hline\hline
 &  Model I   &   Model II    &   Model III \\  \cline{3-4}
  &  & $f_{\rm chv}/g_{\rm chv}=0$ & $f_{\rm chv}/g_{\rm chv}=2/3$ \\
\hline
 $b_u$ (fm)  & 0.5 & 0.45 & 0.45 \\
 $m_u$ (MeV) & 313 & 313 & 313 \\
 $m_s$ (MeV) & 470 & 470 & 470 \\
 $g_u^2$     & 0.766 & 0.056 & 0.132 \\
 $g_s^2$     & 0.846 & 0.203 & 0.250 \\
 $g_{\rm ch}$    & 2.621 & 2.621 & 2.621  \\
 $g_{\rm chv}$   &       & 2.351 & 1.973  \\
 $m_\sigma$ (MeV) & 595 & 535 & 547 \\
 $a^c_{uu}$ (MeV/fm) & 87.5 & 75.3 & 66.2 \\
 $a^c_{us}$ (MeV/fm) & 100.8 & 123.0 & 106.9 \\
 $a^c_{ss}$ (MeV/fm) & 152.2 & 226.0 & 196.7 \\
 $a^{c0}_{uu}$ (MeV)  & $-$77.4 & $-$99.3 & $-$86.6 \\
 $a^{c0}_{us}$ (MeV)  & $-$72.9 & $-$127.9 & $-$109.6 \\
 $a^{c0}_{ss}$ (MeV)  & $-$83.3 & $-$174.20 & $-$148.7 \\
\hline\hline
\end{tabular*}
\end{center}
\end{table}}

The harmonic-oscillator width parameter
$b_u$ is taken to be $0.50$ fm in our original constituent quark
model, and when the vector boson exchanges are included, $b_u$ is
taken to be $0.45$ fm. This means that the bare radius of baryon
becomes smaller when more meson clouds are included in the model,
which sounds reasonable in the sense of the physical picture. The up
(down) quark mass $m_{u(d)}$ and the strange quark mass $m_s$ are
taken to be the usual values: $m_{u(d)}=313$ MeV and $m_s=470$ MeV.
The coupling constant for scalar and pseudoscalar meson field
coupling, $g_{ch}$, is determined according to the relation
\begin{eqnarray}
\frac{g^{2}_{\rm ch}}{4\pi} = \left( \frac{3}{5} \right)^{2}
\frac{g^{2}_{NN\pi}}{4\pi} \frac{m^{2}_{u}}{M^{2}_{N}},
\end{eqnarray}
with the empirical value $g^{2}_{NN\pi}/4\pi=13.67$. The coupling
constant for vector coupling of the vector-meson field is taken to
be $g_{\rm chv}=2.351$, the same as used in the $NN$ case
\cite{lrdai03}. The masses of the mesons are taken to be the
experimental values, except for $\sigma$ meson. The $m_{\sigma}$ is
adjusted to fit the binding energy of the deuteron. The OGE coupling
constants and the strengths of the confinement potential are
determined by baryon masses and their stability conditions. All the
parameters are tabulated in Table I, where the first set is for our
original constituent quark model, the second and third sets are for
the models with vector meson exchanges included by taking $f_{\rm
chv}/g_{\rm chv}$ as $0$ and $2/3$, respectively. Here $f_{\rm chv}$
is the coupling constant for tensor coupling of the vector meson
fields.

From Table I one can see that in models II and III, $g_u^2$ and
$g_s^2$ are much smaller than the values in model I. This means that
when the coupling of quarks and vector meson fields is included in
the non-relativistic constituent quark model, the coupling constants
of OGE will be greatly reduced. Thus the OGE that plays an important
role of the quark-quark short-range interaction in our original
constituent quark model is now nearly replaced by the vector-meson
exchange. In other words, the mechanisms of the quark-quark
short-range interactions in these models are quite different.

We'd like to mention that our previous work concentrated on the
hadron-hadron interactions and it can be strictly proved that
different forms of confinement potential does not make any visible
influence on the theoretical results since the two hadrons are
treated as two color-singlet clusters. In this work we adopt a color
linear confinement potential to study the $ud\bar s\bar s$
one-cluster system, and the method of parameters determination is
the same as in our previous work. Naturally, the three sets of
parameters in Table I still can satisfactorily describe the energies
of the baryon ground states, the binding energy of the deuteron, the
nucleon-nucleon scattering phase shifts and the hyperon-nucleon
cross sections.

\subsection{$ud\bar s\bar s$ configurations}

According to the Pauli principle in each quark (antiquark) pair, in
the spatial symmetry case only six configurations are permitted for
the $ud\bar s\bar s$ system:
\begin{eqnarray*}
I=1, J^P=0^+ & \Longrightarrow & \left\{
\begin{array}{l}
|1\rangle=\left(\{ud\}_1^{\bar 3}\{\bar s\bar s\}_1^3\right)_0 \\
|2\rangle=\left(\{ud\}_0^6\{\bar s\bar s\}_0^{\bar 6}\right)_0
\end{array}
\right. \\
I=0, J^P=1^+ & \Longrightarrow & \left\{
\begin{array}{l}
|3\rangle=\left([ud]_0^{\bar 3}\{\bar s\bar s\}_1^3\right)_1 \\
|4\rangle=\left([ud]_1^6\{\bar s\bar s\}_0^{\bar 6}\right)_1
\end{array}
\right. \\
I=1, J^P=1^+ & \Longrightarrow & \left.
\begin{array}{l}
|5\rangle=\left(\{ud\}_1^{\bar3}\{\bar s\bar s\}_1^3\right)_1
\end{array}
\right. \\
I=1, J^P=2^+ & \Longrightarrow & \left.
\begin{array}{l}
|6\rangle=\left(\{ud\}_1^{\bar 3}\{\bar s\bar s\}_1^3\right)_2
\end{array}
\right.
\end{eqnarray*}

The $s$ quark has isospin-zero so the total isospin of the $ud\bar
s\bar s$ configuration state is determined by the isospin of the
$u,d$ quarks. In the above expressions, \{  \} and [ \, ] represent
the flavor symmetry and antisymmetry, respectively, and the
superscript is the representation of the color SU(3) group, the
subscript is the spin quantum number. Making a re-coupling
calculation, we can express the above six states into two
quark-antiquark pairs, including two color octet $q \bar q$ pairs
and two color singlet $q \bar q$ pairs with $KK$, $KK^*$ and
$K^*K^*$ quantum numbers. The corresponding expressions are given as
follows:
\begin{eqnarray}
|1\rangle\equiv\left(\{ud\}_1^{\bar 3}\{\bar s\bar s\}_1^3\right)_0
=\frac{1}{2}\left((u\bar s)_0^{1}(d\bar
s)_0^{1}\right)_0^{1}-\sqrt{\frac{1}{12}}\left((u\bar
s)_1^{1}(d\bar s)_1^{1}\right)_0^{1} \nonumber\\
-\sqrt{\frac{1}{2}}\left((u\bar s)_0^{8}(d\bar
s)_0^{8}\right)_0^{1}+\sqrt{\frac{1}{6}}\left((u\bar s)_1^{8}(d\bar
s)_1^{8}\right)_0^{1},
\end{eqnarray}
\begin{eqnarray}
|2\rangle\equiv\left(\{ud\}_0^6\{\bar s\bar s\}_0^{\bar 6}\right)_0
=\sqrt{\frac{1}{6}}\left((u\bar s)_0^{1}(d\bar s)_0^{1}\right)_0^{1}
+\sqrt{\frac{1}{2}}\left((u\bar
s)_1^{1}(d\bar s)_1^{1}\right)_0^{1} \nonumber \\
+\sqrt{\frac{1}{12}}\left((u\bar s)_0^{8}(d\bar
s)_0^{8}\right)_0^{1} +\frac{1}{2}\left((u\bar s)_1^{8}(d\bar
s)_1^{8}\right)_0^{1},
\end{eqnarray}
\begin{eqnarray}
|3\rangle\equiv\left([ud]_0^{\bar 3}\{\bar s\bar s\}_1^3\right)_1
=\sqrt{\frac{1}{12}}\left((u\bar s)_0^{1}(d\bar
s)_1^{1}\right)_1^{1} -\sqrt{\frac{1}{12}}\left((u\bar
s)_1^{1}(d\bar s)_0^{1}\right)_1^{1}
+ \sqrt{\frac{1}{6}} \left((u\bar s)_1^{1}(d\bar s)_1^{1}\right)_1^{1} \nonumber\\
-\sqrt{\frac{1}{6}}\left((u\bar s)_0^{8}(d\bar s)_1^{8}\right)_1^{1}
+\sqrt{\frac{1}{6}}\left((u\bar s)_1^{8}(d\bar s)_0^{8}\right)_1^{1}
-\sqrt{\frac{1}{3}}\left((u\bar s)_1^{8}(d\bar
s)_1^{8}\right)_1^{1},
\end{eqnarray}
\begin{eqnarray}
|4\rangle\equiv\left([ud]_1^6\{\bar s\bar s\}_0^{\bar 6}\right)_1
=-\sqrt{\frac{1}{6}}\left((u\bar s)_0^{1}(d\bar
s)_1^{1}\right)_1^{1} +\sqrt{\frac{1}{6}}\left((u\bar
s)_1^{1}(d\bar s)_0^{1}\right)_1^{1}
+\sqrt{\frac{1}{3}}\left((u\bar s)_1^{1}(d\bar s)_1^{1}\right)_1^{1} \nonumber\\
-\sqrt{\frac{1}{12}}\left((u\bar s)_0^{8}(d\bar
s)_1^{8}\right)_1^{1} +\sqrt{\frac{1}{12}}\left((u\bar
s)_1^{8}(d\bar s)_0^{8}\right)_1^{1} +\sqrt{\frac{1}{6}}\left((u\bar
s)_1^{8}(d\bar s)_1^{8}\right)_1^{1},
\end{eqnarray}
\begin{eqnarray}
|5\rangle\equiv\left(\{ud\}_1^{\bar3}\{\bar s\bar s\}_1^3\right)_1
=\sqrt{\frac{1}{6}}\left((u\bar s)_0^{1}(d\bar s)_1^{1}\right)_1^{1}
+\sqrt{\frac{1}{6}}\left((u\bar
s)_1^{1}(d\bar s)_0^{1}\right)_1^{1} \nonumber\\
-\sqrt{\frac{1}{3}}\left((u\bar s)_0^{8}(d\bar
s)_1^{8}\right)_1^{1}-\sqrt{\frac{1}{3}}\left((u\bar s)_1^{8}(d\bar
s)_0^{8}\right)_1^{1},
\end{eqnarray}
\begin{eqnarray}
|6\rangle\equiv\left(\{ud\}_1^{\bar 3}\{\bar s\bar s\}_1^3\right)_2
=\sqrt{\frac{1}{3}}\left((u\bar s)_1^{1}(d\bar s)_1^{1}\right)_2^{1}
-\sqrt{\frac{2}{3}}\left((u\bar s)_1^{8}(d\bar
s)_1^{8}\right)_2^{1},
\end{eqnarray}
where $(u\bar s)(d\bar s)$ represents $\sqrt{1/2}\left[(u\bar
s)(d\bar s)+(d\bar s)(u\bar s)\right)]$ for Eqs. (14), (15), (18)
and (19), and denotes $\sqrt{1/2}\left[(u\bar s)(d\bar s)-(d\bar
s)(u\bar s)\right]$ for Eqs. (16) and (17).

In the actual calculation, the configuration mixing between the
states $|1\rangle$ and $|2\rangle$, as well as $|3\rangle$ and
$|4\rangle$ is considered since these states have same quantum
numbers.

\section{Results and discussions}

{\small
\begin{table}[htb]
\caption{\label{para} Energies (in MeV) of $ud\bar s\bar s$ six
configurations in our constituent quark models (without
configuration mixing).}
\begin{center}
\begin{tabular*}{160mm}{@{\extracolsep\fill}ccccc}
\hline\hline
Configurations  & Model I   &  Model II    &  Model III  & Threshold \\
\hline
 $J^P=0^+$   &    &    &  & \\
 $\left(\{ud\}_1^{\bar 3}\{\bar s\bar s\}_1^3\right)_0$  & 1679 &1650  & 1650 & $KK$~(990) \\
 $\left(\{ud\}_0^6\{\bar s\bar s\}_0^{\bar 6}\right)_0$  & 1828 &1865  & 1833 & $KK$~(990) \\
 $J^P=1^+$   &    &    &  & \\
 $\left([ud]_0^{\bar 3}\{\bar s\bar s\}_1^3\right)_1$  &1698   & 1704 & 1698 & $KK^*$~(1387) \\
 $\left([ud]_1^6\{\bar s\bar s\}_0^{\bar 6}\right)_1$  &1803   & 1812 & 1798 & $KK^*$~(1387) \\
 $\left(\{ud\}_1^{\bar 3}\{\bar s\bar s\}_1^3\right)_1$  &1765   & 1750 & 1742 & $KK^*$~(1387) \\
 $J^P=2^+$   &    &    &  & \\
 $\left(\{ud\}_1^{\bar 3}\{\bar s\bar s\}_1^3\right)_2$  &1904   & 1906  & 1885 & $K^*K^*$~(1784)\\
\hline\hline
\end{tabular*}
\end{center}
\end{table}}

We calculate the energies for six low configurations of $ud\bar
s\bar s$ system in our non-relativistic constituent quark models.
The model parameters we used are shown in Table I, which can
reproduce the NN phase shifts reasonably. As mentioned above, the
parameters of the $s$-channel interactions are fixed by fitting the
masses of $K$ and $K^*$. For reducing the input parameters, here the
size parameter for $K$ and $K^*$ is taken to be the same as that for
baryons. The calculated results (without configuration mixing) are
given in Table II. From this Table, we can see that the results from
model I, II and III are quite similar, although the mechanisms of
the quark-quark short-range interactions are different in these
models, i.e., one is from OGE and the two others are from vector
meson exchanges. This means that the OGE and the vector-meson
exchange can give similar contributions in the $ud\bar s\bar s$
system, the same situation as in the nucleon-nucleon and
kaon-nucleon systems \cite{lrdai03,fhuang05kne}.

In Ref. \cite{cui06}, using the color-magnetic interaction
Hamiltonian with SU(3) flavor symmetry breaking, Cui {\it et al.}
argued that the strong attractive color-magnetic interaction can
reduce the energies of the $ud\bar s\bar s$ systems, and they found
a $I=0$ and $J^P=1^+$ $ud\bar s\bar s$ tetraquark state with a mass
around 1347 MeV. In our constituent quark model calculations, in
model I, the color-magnetic interactions are attractive in both the
isovector $J^P=0^+$ $\left(\{ud\}^{\bar 3}_1\{{\bar s\bar
s}\}^3_1\right)_0$ state and the isoscalar $J^P=1^+$
$\left([ud]_0^{\bar 3}\{\bar s\bar s\}_1^3\right)_1$ state, while
they are repulsive in the other four configurations, and in models
II and III, the OGE is largely reduced and the color-magnetic
attractions are almost replaced by $\rho$ exchange. Furthermore, in
all the models, the $\sigma$ and $\pi$ exchanges provide more
attractive interactions in the isovector $J^P=0^+$
$\left(\{ud\}^{\bar 3}_1\{{\bar s\bar s}\}^3_1\right)_0$ state and
the isoscalar $J^P=1^+$ $\left([ud]_0^{\bar 3}\{\bar s\bar
s\}_1^3\right)_1$ state than in the other four configurations. Thus
the energies of the $\left(\{ud\}^{\bar 3}_1\{{\bar s\bar
s}\}^3_1\right)_0$ and $\left([ud]_0^{\bar 3}\{\bar s\bar
s\}_1^3\right)_1$ states are respectively the lowest one in
$J^P=0^+$ and $J^P=1^+$ cases in various models. However, due to the
high kinetic energies, the attractive interactions are not strong
enough to reduce the energies of these two states to be lower than
the corresponding meson-meson thresholds (see Table II).

The states $\left(\{ud\}_1^{\bar 3}\{\bar s\bar s\}_1^3\right)_0$
and $\left(\{ud\}_0^6\{\bar s\bar s\}_0^{\bar 6}\right)_0$ in
$J^P=0^+$ case, as well as $\left([ud]_0^{\bar 3}\{\bar s\bar
s\}_1^3\right)_1$ and $\left([ud]_1^6\{\bar s\bar s\}_0^{\bar
6}\right)_1$ in $J^P=1^+$ case have the same quantum numbers. Thus
the configuration mixing between them has to be considered. The
results are shown in Table III. Comparing it with Table II, one can
see that the configuration mixing effect is significant, it can
shift the energies over 80 MeV for most configurations. In the
isovector $J^P=0^+$ case, now the energy of the lowest state is
reduced about 170$-$290 MeV in various models, and in the isoscalar
$J^P=1^+$ case, it is about 80$-$130 MeV. Even though, the energies
of these configurations are still higher than their corresponding
meson-meson thresholds. In other words, the stable $ud\bar s\bar s$
tetraquark state cannot yet be obtained.

{\small
\begin{table}[htb]
\caption{\label{para} Energies (in MeV) of $ud\bar s\bar s$ states
with configuration mixing considered.}
\begin{center}
\begin{tabular*}{160mm}{@{\extracolsep\fill}cccc}
\hline\hline
  &  Model I   &   Model II    &   Model III \\
\hline
 $I=1,J^P=0^+$   &1394    & 1483   & 1482  \\
                 &1963    &1952  & 1924 \\
 $I=0,J^P=1^+$   &1567    & 1619   &1615   \\
                 &1887   & 1878 & 1861 \\
 $I=1,J^P=1^+$   &1765   & 1750 & 1742 \\
 $I=1,J^P=2^+$   &1904   & 1906  & 1885 \\
\hline\hline
\end{tabular*}
\end{center}
\end{table}}

The confinement potential is phenomenological, and usually it is
taken as linear, quadratic or error function form. Here we consider
these three various forms of the confinement potential to see the
corresponding effects. The results, from our original
non-relativistic constituent quark model, are shown in Table IV. In
this Table, $r$, $r^2$ and erf represent the confinement potential
adopted as linear, quadratic and error function form, respectively.
We can see that the energies with the confinement potential taken to
be the error function form are always the lowest. But the difference
of various confinement potentials is less than about 30 MeV, which
means different form of the confinement potential has no significant
effect on the energy of the $ud\bar s\bar s$ state.

{\small
\begin{table}[htb]
\caption{\label{para} Energies (in MeV) of the $ud\bar s\bar s$
states in our original constituent quark model with configuration
mixing and three different forms (linear, quadratic and error
function) of the confinement potential considered.}
\begin{center}
\begin{tabular*}{160mm}{@{\extracolsep\fill}lccc}
\hline\hline
  & $r$ & $r^2$ & {\rm erf}  \\  \hline
 $I=1,J^P=0^+$   & 1394    & 1402   & 1390  \\
                 & 1963 & 1996  & 1945 \\
 $I=0,J^P=1^+$   & 1567    & 1568   & 1567   \\
                 & 1887   & 1907 & 1876 \\
 $I=1,J^P=1^+$   & 1765    & 1770   &1763   \\
 $I=1,J^P=2^+$   & 1904    & 1920   &1895 \\
\hline\hline
\end{tabular*}
\end{center}
\end{table}}

{\small
\begin{table}[h!]
\caption{\label{para} Energies (in MeV) of $ud\bar s\bar s$ mixing
states without and with $s$-channel interaction in different models.
}
\begin{center}
\begin{tabular*}{160mm}{@{\extracolsep\fill}lcccccc} \hline \hline
 & \multicolumn{3}{c}{without $s$-channel interaction} & \multicolumn{3}{c}{with $s$-channel interaction}
 \\ \cline{2-4}\cline{5-7}
 & I & II & III & I & II & III \\ \hline
 $I=1,J^P=0^+$  & 1699 & 1910 & 1890 & 1394 & 1483 & 1482  \\
                & 2041 & 2001 & 1991 & 1963 & 1952 & 1924  \\
 $I=0,J^P=1^+$  & 1749 & 1825 & 1828 & 1567 & 1619 & 1615  \\
                & 1984 & 1964 & 1960 & 1887 & 1878 & 1861  \\
 $I=1,J^P=1^+$  & 1876 & 1922 & 1912 & 1765 & 1750 & 1742  \\
 $I=1,J^P=2^+$  & 1964 & 1939 & 1935 & 1904 & 1906 & 1885  \\
\hline \hline
\end{tabular*}
\end{center}
\end{table}}

As we know, the $s$-channel quark-antiquark interaction mechanism is
a complicated and unclear problem. In the study of the structure of
the $\Theta^+$ particle \cite{fhuang04penta}, it has been pointed
out that how to treat the $s$-channel interaction reasonably is very
important. In the $ud\bar s\bar s$ system, the effect of the
$s$-channel interaction should also be examined. We completely omit
this interaction to see its influence, and the results are shown in
Table V, compared with those with $s$-channel interaction. One can
see that the $s$-channel interactions offer quite strong attractions
in all these six $ud\bar s\bar s$ configurations and thus can reduce
the energies of these states about several hundreds MeV in all the
models. In this sense, the effect of the $s$-channel interactions is
significant and un-negligible in the $ud\bar s\bar s$ system.

{\small
\begin{table}[htb]
\caption{\label{para} Energies (in MeV) of $ud\bar s\bar s$ states
with the configuration mixing considered and the size parameter of
$K$, $K^*$ taken as 0.4 fm.}
\begin{center}
\begin{tabular*}{160mm}{@{\extracolsep\fill}cccc}
\hline\hline
  &  Model I   &  Model  II    &  Model  III \\
\hline
 $I=1,J^P=0^+$   &1602    & 1573   & 1572  \\
                 & 1857 &1909  & 1882 \\
 $I=0,J^P=1^+$   &1577    & 1623   &1618   \\
                 &1768   & 1833 & 1817 \\
 $I=1,J^P=1^+$   &1771   & 1754 & 1745 \\
 $I=1,J^P=2^+$   &1821   & 1872  & 1852 \\
\hline\hline
\end{tabular*}
\end{center}
\end{table}}

In the above calculations, the size parameter of $K$ and $K^*$
mesons are taken to be the same as the baryons' in order to reduce
the free parameters. It seems more reasonable to choose the size
parameter for the mesons to be smaller than that for the baryons. We
then perform a variational method calculation for the energies of
the $ud\bar s\bar s$ states with the size parameter of $K$ and $K^*$
taken to be $0.4$ fm, and the results are shown in Table VI. We
notice that one of the $I=0$ $J^P=1^+$ $ud\bar s\bar s$
configurations is a very interesting state in model I. Its energy is
1768 MeV, lower than the threshold of $K^*K^*$, and the
corresponding root mean square radius of this state is about $0.57$
fm. Especially, the structure of this state is very interesting
since it contains very few components of $K K^*$, as can be seen
from the expression of its wave function given below:
\begin{alignat}{4}
|4\rangle' &=&& \,-\,0.17\left((u\bar s)_0^{1}(d\bar
s)_1^{1}\right)_1^{1} && \,+\, 0.17\left((u\bar s)_1^{1}(d\bar
s)_0^{1}\right)_1^{1} && \,+\, 0.71\left((u\bar s)_1^{1}(d\bar
s)_1^{1}\right)_1^{1} \nonumber\\
&&& \,-\, 0.47\left((u\bar s)_0^{8}(d\bar s)_1^{8}\right)_1^{1} &&
\,+\, 0.47\left((u\bar s)_1^{8}(d\bar s)_0^{8}\right)_1^{1} && \,+\,
0.0056\left((u\bar s)_1^{8}(d\bar s)_1^{8}\right)_1^{1}.
\end{alignat}
From this equation it is clear to see that besides 43.8\% part of
two color octet $q \bar q$ pairs, the component of two color singlet
$q\bar q$ pairs is 50.4\% for $K^*K^*$ and only 5.8\% for $KK^*$,
which means the $K^*K^*$ component is dominate and comparatively the
$KK^*$ component is very small in the color-singlet $q\bar q$-$q\bar
q$ part, thus the $I=0$ $J^P=1^+$ $ud\bar s\bar s$ configuration
might have a few possibility decaying into $K$ and $K^*$.

\begin{figure}[htb]
\epsfig{file=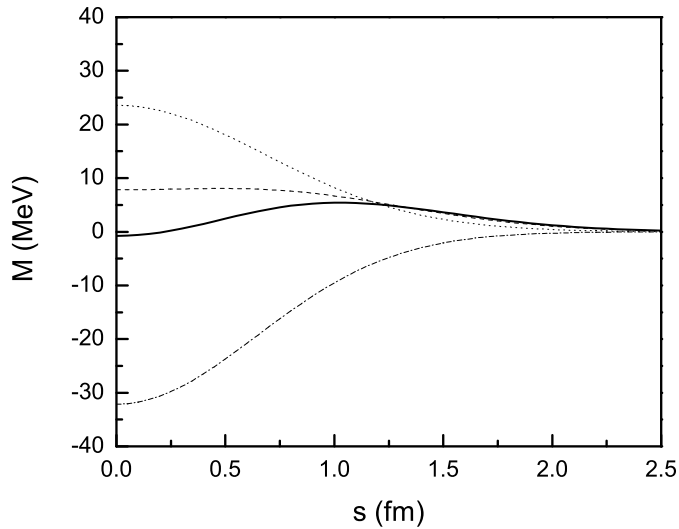,width=10cm} \caption{\small The matrix
elements of the interaction potential. The solid line represent the
matrix elements of interaction potential between the $|4\rangle'$
state and the $KK^*$ channel. The dotted, dash-dotted and dashed
lines represent the contributions of $\left((u\bar s)_0^{1}(d\bar
s)_1^{1}\right)_1^{1}$, $\left((u\bar s)_1^{1}(d\bar
s)_1^{1}\right)_1^{1}$, and the two color-octet components in
$4\rangle'$ state, respectively.}
\end{figure}

Furthermore, in order to see the effect of the coupling between this
$I=0$ $J^P=1^+$ $ud\bar s\bar s$ state and the $KK^*$ channel, we
calculate the interaction matrix elements between the $|4\rangle'$
state and $KK^*$ channel in model I by treating $K$ and $K^*$ as two
clusters with a distance $S$,
\begin{equation}
\langle V \rangle=\langle 4'|\sum_{i\in K \atop j\in K^*}V_{ij}|KK^*
\psi(\vec{R}-\vec{S})\rangle,
\end{equation}
where $\psi(\vec{R}-\vec{S})$ is the relative motion wave function
of the two clusters $K$ and $K^*$, and for simplicity we take it as
\begin{equation}
\psi(\vec{R}-\vec{S})=\left(\omega \mu/\pi\right)^{3/4}{\rm
exp}[-\omega \mu(\vec{R}-\vec{S})^2/2],
\end{equation}
here $S$ is the generator coordinate which can qualitatively
describe the distance between the two clusters. The calculated
results are shown in Figure 1, along with the contributions of
$\left((u\bar s)_0^{1}(d\bar s)_1^{1}\right)_1^{1}$, $\left((u\bar
s)_1^{1}(d\bar s)_1^{1}\right)_1^{1}$ and the two color-octet $q
\bar q$ pairs components in the $|4\rangle'$ state. In Fig. 1, M
denotes the interaction matrix elements, and $S$ can qualitatively
describe the distance between the two clusters $K$ and $K^*$.

From Fig. 1 one can see that the contribution of color-octet
components to the interaction matrix elements is the smallest which
is apparent and can be easily understood, while the $\left((u\bar
s)_0^{1}(d\bar s)_1^{1}\right)_1^{1}$ component also has a small
contribution due to the small component in $|4\rangle'$ (smaller
than $6\%$), although the matrix element $\langle\left((u\bar
s)_0^{1}(d\bar s)_1^{1}\right)_1^{1}|\sum_{i\in K \atop j\in
K^*}V_{ij}|KK^* \psi(\vec{R}-\vec{S})\rangle$ is comparatively big.
The solid line in Fig. 1 clearly tells us that the total
contribution of $4\rangle'$ to the interaction matrix elements, i.e.
$\langle 4'|\sum_{i\in K \atop j\in K^*}V_{ij}|KK^*
\psi(\vec{R}-\vec{S})\rangle$, is very small, less than 6 MeV and
inclines to zero with the increase in distance between $K$ and
$K^*$. Hence, the coupling between this interesting $|4\rangle'$
state and the $KK^*$ channel can be regarded as quite small. This
means the $I=0$ $J^P=1^+$ $ud\bar s\bar s$ state has a few
possibility decaying into two separate $K$ and $K^*$ via interaction
potential. Furthermore, since the energy of this state, 1768 MeV, is
lower than the $K^*K^*$ threshold (1784 MeV), it cannot decay into
$K^*K^*$ final state. This means this state would possibly have a
narrow width, and might be treated as a good candidate for the
$ud\bar s\bar s$ tetraquark state.

As discussed in Ref. \cite{cui06}, it seems worth searching this
state in $K^+d \rightarrow p + p + K^- + \mathcal{T}^+$ or $J/\psi
(\Upsilon) \rightarrow K^- + {\bar K}^0 + \mathcal{T}^+$ channels in
the future experiments [Here $\mathcal{T}^+$ denotes the $ud\bar
s\bar s$ tetraquark state with $I=0$ and $J^P=1^+$]. The
experimental information about the existence of this state will help
us test the validity of the application of our constituent quark
models to the study of multi-quark states, although these models are
quite successful in the investigations of $NN$, $NY$, and $KN$
interactions
\cite{zyzhang97,lrdai03,fhuang05kne,fhuang04kn,fhuang04nkdk}.

\section{Summary}

The structures of $ud\bar s\bar s$ states with $J^P=0^+$, $1^+$, and
$2^+$ are studied in our non-relativistic constituent quark models
with the meson and one gluon exchanges included in the quark-quark
interaction potentials. We calculate the energies of six low-lying
$ud\bar s\bar s$ configurations by use of the variational method.
The configuration mixing between the states with same quantum
numbers are considered. The effect of different forms of color
confinement potentials and the contributions of $s$-channel $q\bar
q$ interactions are also examined. The results show that the
different forms (linear, quadratic, and error function) of the
confinement potential just give similar contributions, and the
$s$-channel interactions can reduce the energy of the $ud\bar s\bar
s$ system several hundred MeV. With the model parameters determined
by the same method as in our previous work, the calculated energies
of all the $ud\bar s\bar s$ configurations are higher than the
corresponding meson-meson thresholds. But when the size parameter
for the mesons is adjusted to be $0.4$ fm, a value smaller than that
for the baryons, the $ud\bar s\bar s$ configuration with $I=0$ and
$J^P=1^+$ is found to lie lower than the $K^*K^*$ threshold,
furthermore, this state has a very small $KK^*$ component and the
interaction matrix elements between this state and $KK^*$ is
comparatively small, thus its coupling to the $KK^*$ channel will
consequently be weak and it might be regarded as a possible
tetraquark candidate. A further dynamical calculation would be done
in the future work.

\begin{acknowledgements}
This work was supported in part by the National Natural Science
Foundation of China (No. 10475087) and China Postdoctoral Science
Foundation (No. 20060400093).
\end{acknowledgements}

\end{document}